\documentclass[
reprint,
amsmath,amssymb,
aps, prl, groupedaddress,
longbibliography
]{revtex4-2}

\usepackage{graphicx}
\usepackage{dcolumn}
\usepackage{bm}

\usepackage[utf8]{inputenc}
\usepackage[T1]{fontenc}
\usepackage{mathptmx}
\DeclareMathAlphabet{\mathcal}{OMS}{cmsy}{m}{n}

\usepackage{braket}
\usepackage{amsmath}
\usepackage{caption}
\usepackage{subcaption}
\usepackage{gensymb}
\usepackage{xcolor}
\usepackage{mwe}
\usepackage{stackengine}

\begin{document}
	
\preprint{AIP/123-QED}

\title[]{Entangling-gate error from coherently displaced motional modes of trapped ions}

\author{B.~P. Ruzic}
\email{bruzic@sandia.gov}
\author{T.~A. Barrick}
\author{J.~D. Hunker}
\author{R.~J. Law}
\author{B.~K. McFarland}
\author{H.~M. McGuinness}
\author{L.~P. Parazzoli}
\author{J.~D. Sterk}
\author{J.~W. Van Der Wall}
\author{D. Stick}
\affiliation{Sandia National Laboratories, Albuquerque, New Mexico 87185, USA}

\date{\today}

\begin{abstract}
Entangling gates in trapped-ion quantum computing have primarily targeted stationary ions with initial motional distributions that are thermal and close to the ground state.
However, future systems will likely incur significant non-thermal excitation due to, e.g., ion transport, longer operational times, and increased spatial extent of the trap array. In this paper, we analyze the impact of such coherent motional excitation on entangling-gate error by performing simulations of M\o lmer-S\o renson (MS) gates on a pair of trapped-ion qubits with both thermal and coherent excitation present in a shared motional mode at the start of the gate. We discover that a small amount of coherent displacement dramatically erodes gate performance in the presence of experimental noise, and we demonstrate that applying only limited control over the phase of the displacement can suppress this error. We then use experimental data from transported ions to analyze the impact of coherent displacement on MS-gate error under realistic conditions.

\end{abstract}

\maketitle

\section{Introduction}
\label{sec:intro}

The ability to achieve precise control of qubits in the presence of noise is fundamental to the progress of quantum computation and quantum sensing.
The M\o lmer-S\o renson (MS) two-qubit entangling gate~\cite{molmer:1999} for trapped-ion quantum computation is a good example of this, as the gate is designed to reduce the error caused by initial ion motion.
While trapped-ion qubits encode quantum information in long-lived internal states, ion motion mediates the interactions between qubits, and noise that affects the quantized motional state can significantly degrade the performance of entangling gates. This use of a noisy degree of freedom to mediate two-qubit interactions is not unique to trapped-ion systems; for instance, entangling gates in neutral-atom systems employ a short-lived Rydberg state for this purpose~\cite{mitra:2020}.
Motional excitation also plays a critical role in quantum-sensing applications, including trapped-ion motional sensors~\cite{mccormick:2019} and inertially-sensitive neutral-atom interferometers, for which motional noise that persists after state preparation is predicted to be one of the dominant error sources~\cite{brif:2020}.

A significant amount of research in trapped-ion quantum computation has focused on reducing the electric-field noise that causes ``anomalous heating''~\cite{bruzewicz:2015, boldin:2018} and degrades MS-gate performance. The gate error results from incoherent excitation of the motional state both during the gate, while it is temporarily entangled with the internal-state qubit, and prior to the gate by corrupting the initial motional state.
For an experiment that performs limited transport tailored to the specific system \cite{wan:2019} or sympathetically cools the ions prior to the gate, reference~\cite{sorensen:2000} accurately predicts the gate error because the ions are close to their motional ground state and because thermal excitation contributes the majority of motion-related error.

This picture grows significantly more complicated
for large, integrated trapped-ion systems that rely on extensive shuttling operations. This is particularly true for the quantum charge coupled device (QCCD) architecture \cite{kielpinski:2002}, where experiments motivated by this concept have demonstrated linear \cite{walther:2012}, split/join \cite{bowler:2012, palmero:2015}, and junction \cite{blakestad:2011, shu:2014} transport in both surface and 3D traps. Motional excitation over the course of an algorithm, whether from persistent voltage noise (i.e. anomalous heating) or transport induced excitation, is especially damaging to entangling-gate performance because each gate is sensitive to the accumulated excitation. A promising mitigation strategy relies on sympathetically cooling the motional degrees of freedom of the qubits while preserving any encoded quantum information~\cite{lin:2013}, but this is costly in both time and infrastructure. Considerable time would be saved if sympathetic cooling were only needed occasionally to reduce small amounts of excitation. 

In the work described here, we seek to better understand the impact of motional excitation on MS gates by computing the gate error that arises from both coherent and thermal excitation in the initial state of the gate-mediating motional mode. Both types of motion can arise from -- or be influenced by -- environmental and control sources, and while they do not affect the internal qubit directly, their accumulated impact prior to the two-qubit gate degrades its performance through temporary spin-motion entanglement. In particular, we investigate how each type of motion differently exacerbates the gate error resulting from fluctuations in trap frequency, a ubiquitous source of experimental noise. 

Imperfections in transport control inevitably lead to some degree of motional excitation, but careful control design can tilt the balance between thermal excitation and coherent displacement. For example, slow transport speeds can result in less coherent displacement after transport but contribute to longer operational times that introduce more anomalous heating. Additionally, background electric fields can drift over minutes and hours, altering the ion trajectory \cite{bowler:2012} and increasing the magnitude of induced coherent displacement over time. By analyzing our simulations of MS-gate error, we find that a small amount of coherent displacement at the start of the gate leads to a large gate error, with a strong dependence on the phase of the coherent displacement. Near the lower limit of experimentally-feasible trap frequency noise, we show that optimizing the phase of a two-quanta coherent displacement reduces the associated gate error by 86\% (52\%) compared to the least optimal phase, as quantified by the process infidelity (diamond distance). We then apply our simulations to experimental data in which the application of a background electric field coherently displaces the motional modes of an ion. 
\section{MS-gate model}
\label{sec:model}

We model the application of an MS gate on two ions that are part of a linear chain of ions in a surface trap using the Hamiltonian,
\begin{equation}
H(t) = -\eta\Omega J_y\left(a e^{i\delta t} + a^\dagger e^{-i\delta t}\right),
\end{equation}
which is in a rotating frame with respect to the atomic and trap degrees of freedom. The collective spin operator $J_y$ has the form: $J_y = (\sigma_{y1}+\sigma_{y2})/2$, where $\sigma_{yj}$ is the $y$ Pauli spin operator for the $j$-th ion targeted by the gate. The Lamb-Dicke parameter $\eta$ is the same for both ions, and $\Omega$ is the Rabi rate of the carrier transition for both ions. The operators $a^\dagger$ and $a$ are the raising and lowering operators, respectively, for a harmonic oscillator that represents a single motional mode of the ion chain with angular frequency~$\nu$. During the gate, a dual-tone laser illuminates the ions with detunings $\pm\delta=\pm(\delta_\text{c}-\nu)$ from their blue and red motional sideband transitions, respectively, where the parameter $\delta_\text{c}$ is the detuning of the blue-detuned laser tone from the carrier transition. For simplicity, we have made the Lamb-Dicke approximation: $e^{i\eta(a+a^\dagger)}\approx 1+i\eta(a+a^\dagger)$. We have also neglected the carrier transition and the far-off-resonant sideband transitions. 

The exact analytic solution for the propagator $U(t)$ is,
\begin{gather}
\label{eq:U}
U(t) = e^{-i\mathcal{B}(t)J_y^2}D(J_y\alpha(t)), \nonumber \\
\mathcal{B}(t)=\int_0^t \left(\text{Im}[\alpha(t')]\frac{\mathrm{d}\text{Re}[\alpha(t')]}{\mathrm{d}t'} -\text{Re}[\alpha(t')]\frac{\mathrm{d}\text{Im}[\alpha(t')]}{\mathrm{d}t'}\right)\mathrm{d}t',
\end{gather}
which is equivalent to the solution in reference~\cite{sorensen:2000}.
The displacement operator $D(J_y\alpha(t))=\exp\left[J_y(\alpha(t) a^\dagger - \alpha^*(t) a)\right]$ is conditioned on the spin state of the targeted ions, and $\alpha(t)$ describes the phase-space trajectory of the ion chain. The phase $\mathcal{B}(t)$, which governs the amount of spin entanglement accrued during the gate, is
positive (negative) for clockwise (counter-clockwise) trajectories. 
In terms of the parameters of $H(t)$,  
\begin{gather}
\alpha(t) = \frac{\eta\Omega}{\delta}\left(1 - e^{-i\delta t}\right), \nonumber \\
\mathcal{B}(t)=\frac{\eta^2\Omega^2}{\delta^2}\left(\delta t-\sin\delta t\right).
\end{gather}

To simulate the MS gate, we use $U(t)$ to propagate the density matrix of the ions $\rho(t)$ from their initial state,
\begin{equation}
\rho(0) = \rho_\text{spin}\otimes\rho_\text{motion},
\end{equation}
where $\rho_\text{spin}$ and $\rho_\text{motion}$ describe the initial spin and motional degrees of freedom, respectively, to the state $\rho(\tau)$ at the end of the gate. The error of this gate depends on the character of the initial motional state $\rho_\text{motion}$, which accumulates all prior motional excitation since the ions were last cooled, including excitation from gates, heating, and transport.

\section{Initial motional state}

The experimental realization of a quantum algorithm on a linear chain of ions can invoke both coherent and incoherent motional excitation, which we represent as a coherent displacement in phase space $\alpha=|\alpha|e^{i\phi}$ and an increase in the ion temperature $T$, respectively. Under this premise, an ion chain cooled to its motional ground state at the start of the algorithm arrives in a thermal mixture of coherently displaced Fock states immediately before an MS gate occurs. We represent the $n$-th Fock state of the harmonic oscillator by $\ket{n}$, and we represent a coherently displaced Fock state by, 
\begin{gather}
\ket{\alpha, n}=D(\alpha)\ket{n},
\end{gather}
where $D(\alpha)=\exp\left(\alpha a^\dagger - \alpha^* a\right)$ is the displacement operator. Hence, we describe the initial motional state for the gate by the partial density matrix,
\begin{gather}
\label{eq:rho_motion}
\rho_\text{motion} = \sum_{n=0}^\infty\frac{1}{1+\bar{n}_\text{th}}\left(\frac{\bar{n}_\text{th}}{1+\bar{n}_\text{th}}\right)^n \ket{\alpha, n}\bra{\alpha, n},
\end{gather}
where $\bar{n}_\text{th}=(\exp(\hbar\nu/k_\text{B}T)-1)^{-1}$ in which $k_\text{B}$ is the Boltzmann constant.

The coherently displaced Fock state $\ket{\alpha, n}$ has the following expansion onto Fock states \cite{wunsche:1991},
\begin{gather}
\ket{\alpha, n} = \sum_{m=0}^\infty C_m^{(\alpha, n)} \ket{m} , \nonumber \\
\label{eq:alpha_n}
C_m^{(\alpha, n)} = e^{-|\alpha|^2/2}\sqrt{n!/m!}\alpha^{m-n}L^{(m-n)}_n(|\alpha|^2),
\end{gather}
where $L^{(m-n)}_n$ is the generalized $n$-th order Laguerre polynomial. The expectation value of the number operator $\hat{n}=a^\dagger a$ in the state $\rho_\text{motion}$ is $\braket{\hat{n}}=|\alpha|^2 + \bar{n}_\text{th}$, and this quantity determines the average motional energy $\hbar\nu(1/2+\braket{\hat{n}})$. Even though $\braket{\hat{n}}$ contains equal contributions from $|\alpha|^2$ and $\bar{n}_\text{th}$, coherent displacement generates correlations between different Fock states and, in this way, produces a fundamentally different motional state than thermal excitation.

\begin{figure}[t]
\raggedright
\begin{subfigure}[b]{0.47\textwidth}
\raggedright
\resizebox{1.0\textwidth}{!}{
\includegraphics{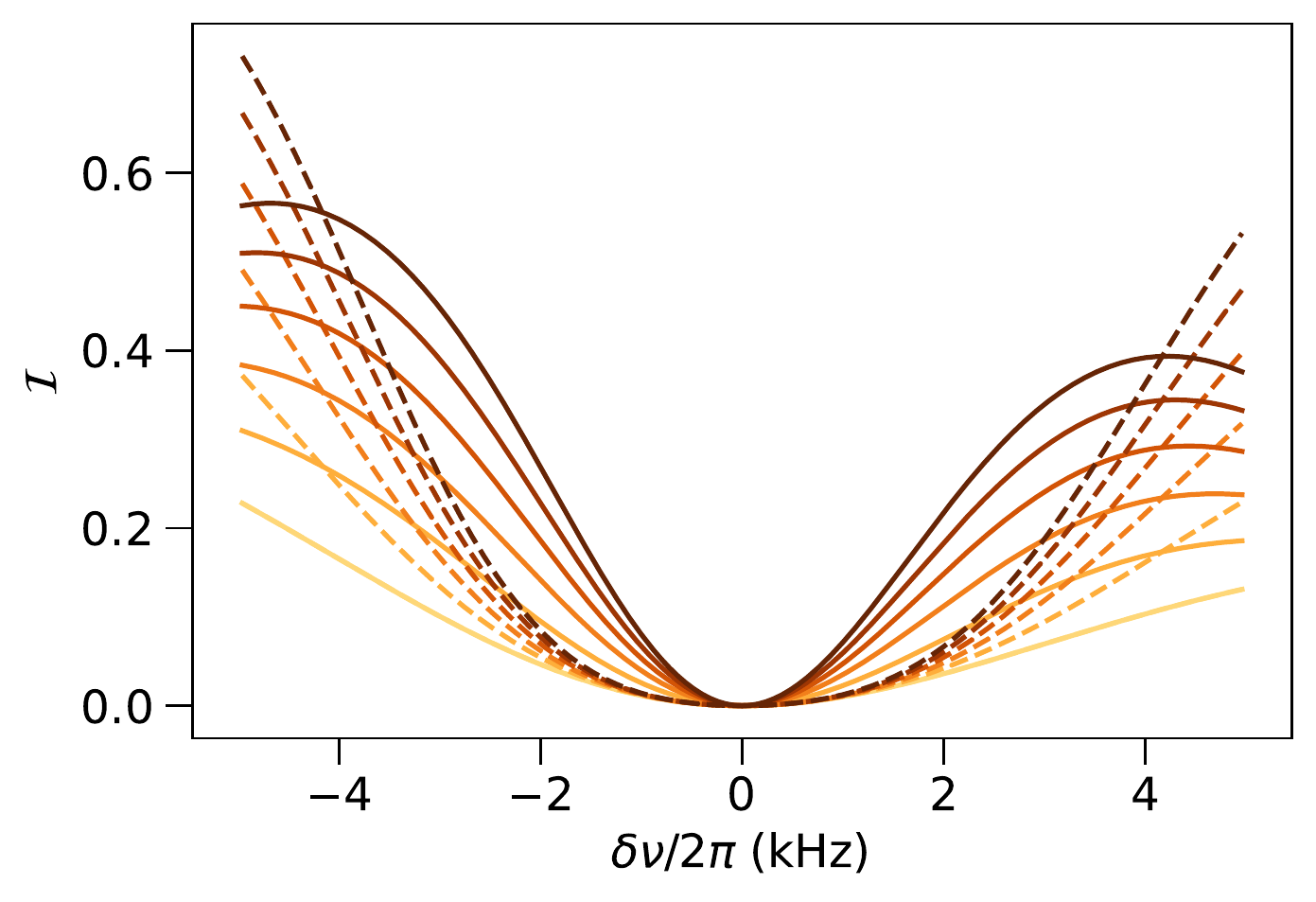}}
\end{subfigure}
\hfill
\begin{subfigure}[b]{0.47\textwidth}
\raggedright
\resizebox{1.0\textwidth}{!}{
\includegraphics{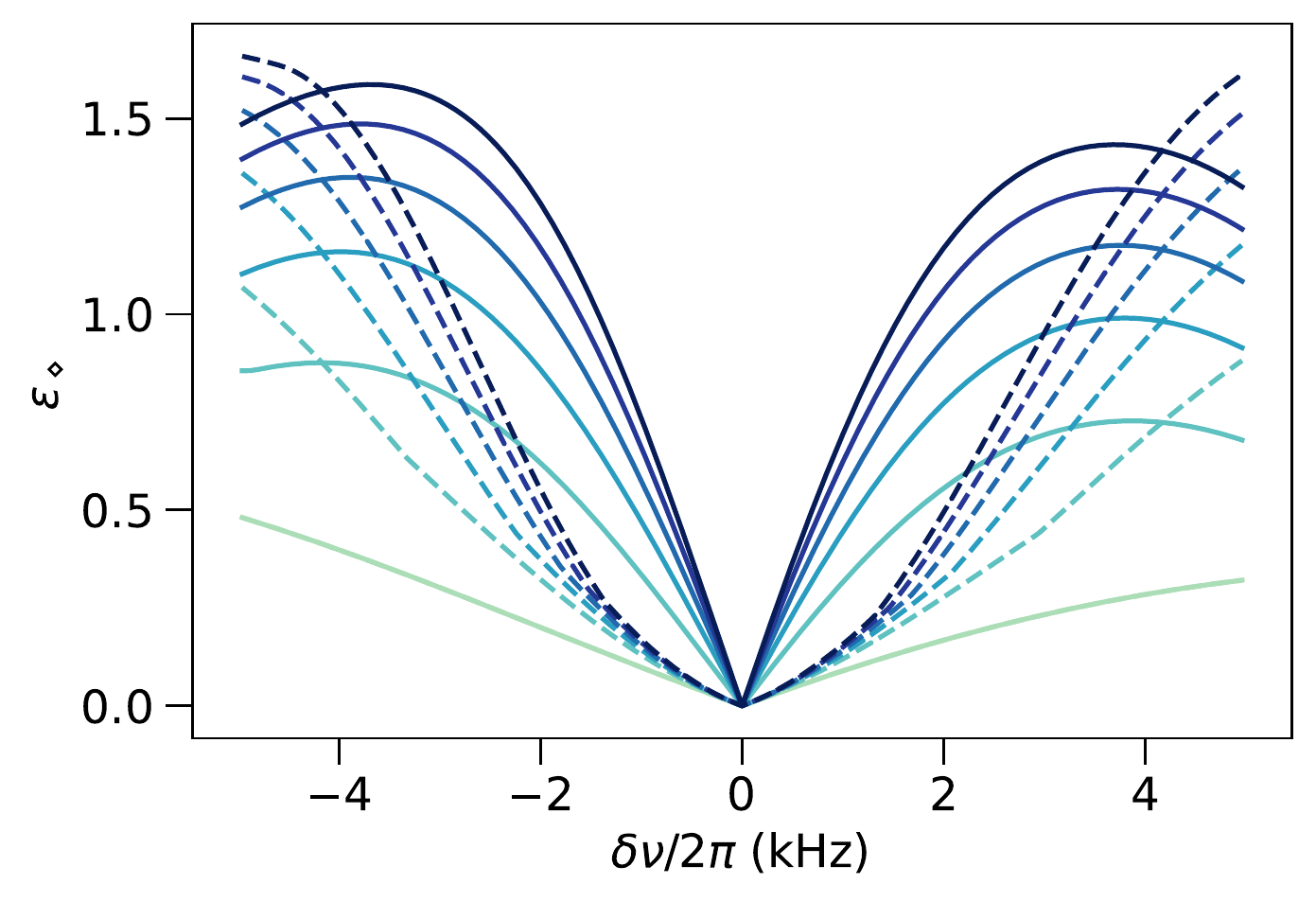}}
\end{subfigure}
\caption{\raggedright Infidelity $\mathcal{I}$ and diamond distance $\epsilon_\diamond$ for an MS gate vs.\ trap frequency error $\delta\nu/2\pi$ with $\bar{n}_\text{th}=0$. From bottom (lightest) to top (darkest) in each plot, the calculations are for $|\alpha|^2=0$ to 2 in steps of $0.4$ with $\phi=0$ (solid lines) and with $\phi=\pi/2$ (dashed lines). For $|\alpha|^2=0$, the gate error is the same for both values of $\phi$.}
\label{fig:fid_vs_trap_freq}
\end{figure}

\section{MS-gate error}
\label{sec:error_rate}

We quantify the error of the MS gate by computing its infidelity $\mathcal{I}$ and diamond distance $\epsilon_\diamond$ according to references~\cite{wallman:2016, nielsen:2020}.
While the initial motional-state distribution affects the gate error associated with multiple control errors, we focus on the interplay between this distribution and trap frequency fluctuations because of their relatively large impact on gate performance compared to other control errors (e.g. laser intensity drift) and because the same physical noise sources (e.g. background electric fields and imperfect control voltages) contribute to both quantities. We also consider different values of the initial phase $\phi$ of the coherent state prior to the gate.

Although our MS-gate model is appropriate for a wide range of experimental conditions, we provide a concrete example by simulating an MS gate designed to complete $K=2$ counter-clockwise loops in phase space during a gate duration of $\tau=60$~$\mu$s. This requires $\delta/2\pi=-K/\tau=-33.3$~kHz to close the loops and $\eta\Omega/2\pi=\sqrt{K}/2\tau=11.8$~kHz to produce $\mathcal{B}(\tau)=-\pi/2$. We also choose the motional frequency $\nu/2\pi=3$~MHz, which is a representative value for the axial, center-of-mass motional mode of a linear chain of $^{40}$Ca$^+$ ions in a surface trap. For this mode, $\nu/2\pi$ is equal to the axial trap frequency. 

We incorporate trap frequency error into the MS-gate model by shifting $\nu$: $\nu = \nu_0 + \delta\nu$, where $\nu_0/2\pi=3$~MHz. This causes $\delta$ to deviate from its optimal value: $\delta=\delta_0-\delta\nu$, where $\delta_0/2\pi=-33.3$~kHz, while $\tau$ and $\eta\Omega$ remain fixed. Fig.~\ref{fig:fid_vs_trap_freq} shows how the simulated MS-gate infidelity $\mathcal{I}$ and the diamond distance $\epsilon_\diamond$ depend on the trap frequency error $\delta\nu/2\pi$ for several values of $|\alpha|^2$ with $\bar{n}_\text{th}=0$, for both $\phi=0$ and $\phi=\pi/2$.
For $|\delta\nu|/2\pi\lesssim 3$~kHz, the gate error grows as the magnitude of trap frequency error increases. For $|\delta\nu|/2\pi\gtrsim 3$~kHz, the gate error oscillates and remains large. An experiment would observe these features if the trap frequency drifts away from $\nu_0/2\pi$ over the course of many experiments. The gate error is more sensitive to $\delta\nu$ for higher values of $|\alpha|^2$, and this sensitivity depends on $\phi$. Hence, for a certain acceptable gate error, the values of $\phi$ and $|\alpha|^2$ set the time between necessary re-calibrations of $\nu$.

As shown in Fig.~\ref{fig:fid_vs_trap_freq}, significant trap frequency error ($|\delta\nu|/2\pi=3$ to 5 kHz) and only modest coherent displacement ($|\alpha|^2=0.4$ to 2) generates a large gate error that is comparable for both $\phi=0$ and $\phi=\pi/2$.
However, for values of $\delta\nu$ that produce experimentally relevant gate errors, coherent states with $\phi=\pi/2$ show a significant reduction in the sensitivity of gate error to $\delta\nu$, as compared to $\phi=0$. For example, with $\delta\nu/2\pi=-600$ Hz and $|\alpha|^2 = 2$, $\mathcal{I}=0.030$ (0.0045) and $\epsilon_\diamond=0.45$ (0.084) for $\phi=0$ $(\pi/2)$.

\begin{figure}[t]
\raggedright
\begin{subfigure}[b]{0.239\textwidth}
\raggedright
\resizebox{1.0\textwidth}{!}{
\includegraphics{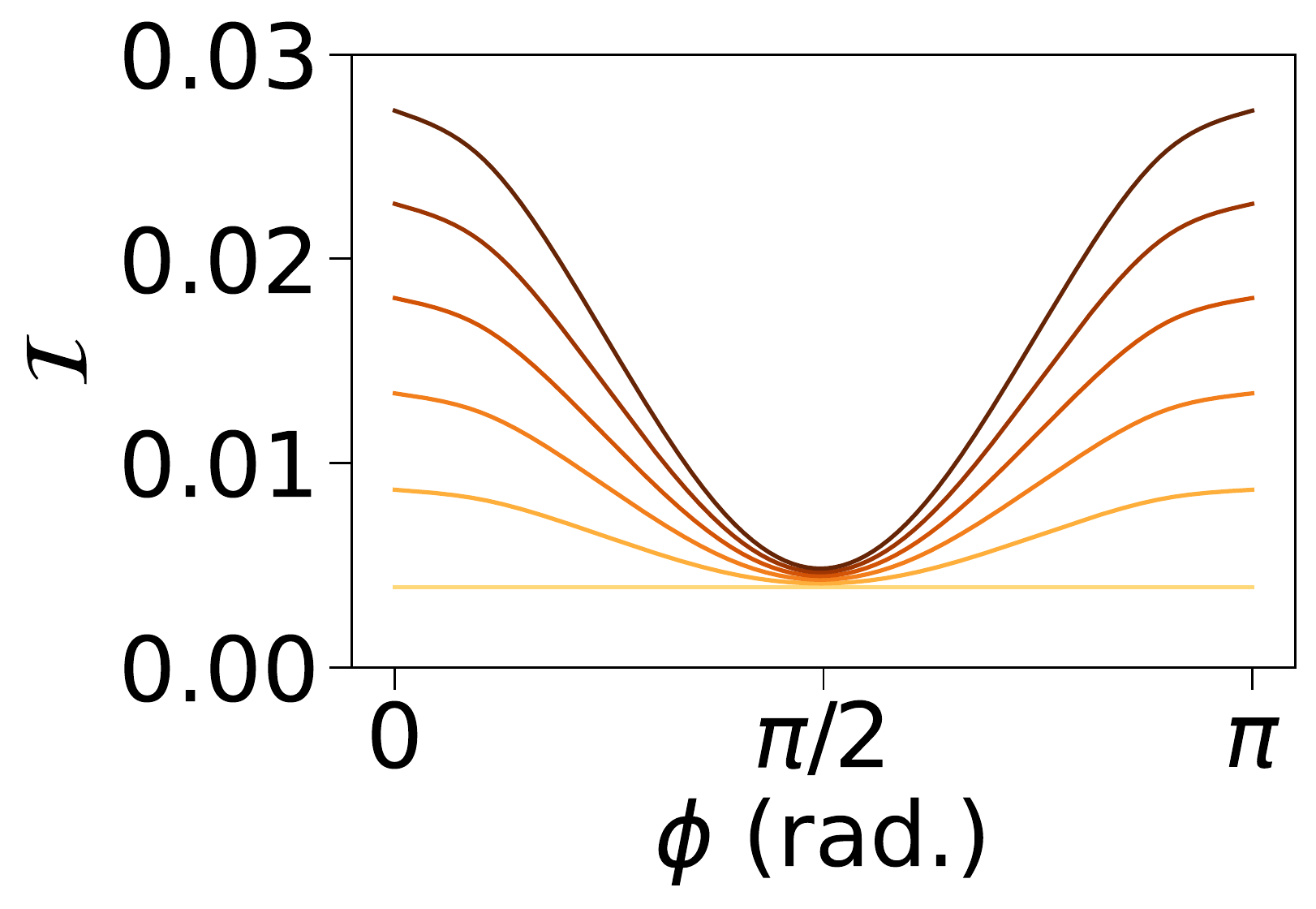}}
\end{subfigure}
\begin{subfigure}[b]{0.239\textwidth}
\raggedright
\resizebox{1.0\textwidth}{!}{
\includegraphics{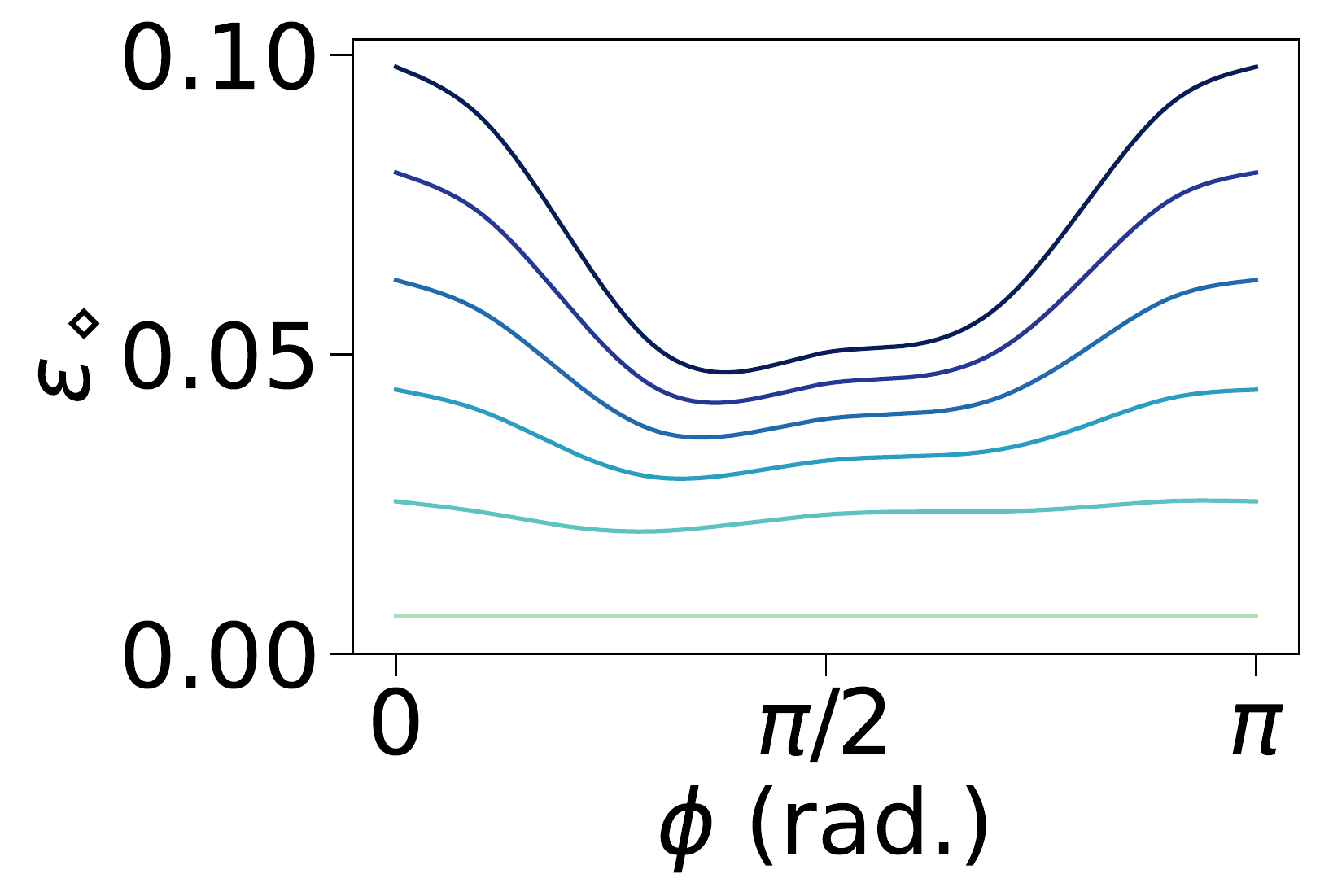}}
\end{subfigure}
\caption{\raggedright Infidelity $\mathcal{I}$ and diamond distance $\epsilon_\diamond$ for an MS gate averaged over a Gaussian distribution of trap frequencies with width $\sigma/2\pi = 600$ Hz and centered at $\nu_0/2\pi=3$ MHz vs.\ the phase $\phi$ of the initial coherent state with $\bar{n}_\text{th}=0$. From bottom (lightest) to top (darkest) in each plot, the calculations are for $|\alpha|^2=0$ to 2 in steps of $0.4$.}
\label{fig:fid_vs_phi}
\end{figure}

In addition to drifting over the course of many experiments, the trap frequency fluctuates from shot to shot during a single experiment due to voltage noise on the electrodes and other sources. We model this kind of noise by averaging the MS gate over a Gaussian distribution of trap frequencies with width $\sigma/2\pi$ and centered at $\nu_0/2\pi=3$ MHz. Fig.~\ref{fig:fid_vs_phi} shows how the infidelity $\mathcal{I}$ and the diamond distance $\epsilon_\diamond$ of the average MS gate depend on $\phi$ when $\sigma/2\pi=600$ Hz. This value of $\sigma/2\pi$ is representative of trap frequency fluctuations in modern surface traps. Initial states with phase $\phi + k \pi$ have the same gate error, where $k$ is any integer.


From Fig.~\ref{fig:fid_vs_phi} we see that $\phi=\pi/2$ provides maximum robustness to trap frequency noise when $\sigma/2\pi=600$ Hz. The process infidelity $\mathcal{I}$ has a dramatic minimum at $\phi=\pi/2$. Although the minimum of $\epsilon_\diamond$ is not at exactly $\phi=\pi/2$, this value of $\phi$ is in the center of an approximately flat region of $\epsilon_\diamond$ vs.\ $\phi$, and $\phi$ itself can be altered by trap frequency noise. In this figure, with $\sigma/2\pi=600$ Hz and $|\alpha|^2=2$, $\mathcal{I} = 0.027$ (0.0048) and $\epsilon_\diamond = 0.098$ (0.050) at $\phi=0$ $(\pi/2)$. This corresponds to an 82\% reduction in $\mathcal{I}$ and a 49\% reduction in $\epsilon_\diamond$ by changing $\phi$ from $0$ to $\pi/2$.
The minimum in the gate error at $\phi=\pi/2$ becomes more pronounced for smaller values of $\sigma$. For example, when $\sigma/2\pi=200$ Hz, changing $\phi$ from 0 to $\pi/2$ corresponds to an 86\% reduction in $\mathcal{I}$ and a 52\% reduction in $\epsilon_\diamond$.
For $\sigma/2\pi\gtrsim 600$ Hz, the minimum $\mathcal{I}$ remains at $\phi=\pi/2$, and two minima emerge in $\epsilon_\diamond$ vs.\ $\phi$.


The sensitivity of transport to experimental conditions like electrode voltages, filters, and relative timing of pulses prevents the calculation of $\phi$ \emph{a priori}. However, providing a time delay after transport can vary the value of $\phi$ at the start of the gate, and minimizing $\mathcal{I}$ vs.\ the time delay can select $\phi=\pi/2$. This procedure will simultaneously minimize $\epsilon_\diamond$ and optimize the performance of quantum algorithms that use these gates.

Fig.~\ref{fig:fid_vs_phi} shows that $\epsilon_\diamond$ is more sensitive than $\mathcal{I}$ to coherent displacement. This is consistent with coherent displacement causing a substantial amount of coherent gate error, as opposed to the purely stochastic error caused by thermal excitation. As quantum circuits amplify coherent gate error, it is especially damaging to long quantum algorithms that involve many gates~\cite{iverson:2020}. As a result, the balance between coherent displacement and thermal excitation plays a critical role in the design of transport solutions that maximize circuit performance, including the choice of transport speeds. Although this study focuses on trap frequency noise, other noise sources (e.g. uncontrolled ac-Stark shifts) may amplify the detrimental effect of coherent displacement on high-fidelity gates. 

\begin{figure}[t]
\raggedright
\begin{subfigure}[b]{0.239\textwidth}
\raggedright
\resizebox{1.0\textwidth}{!}{
\includegraphics{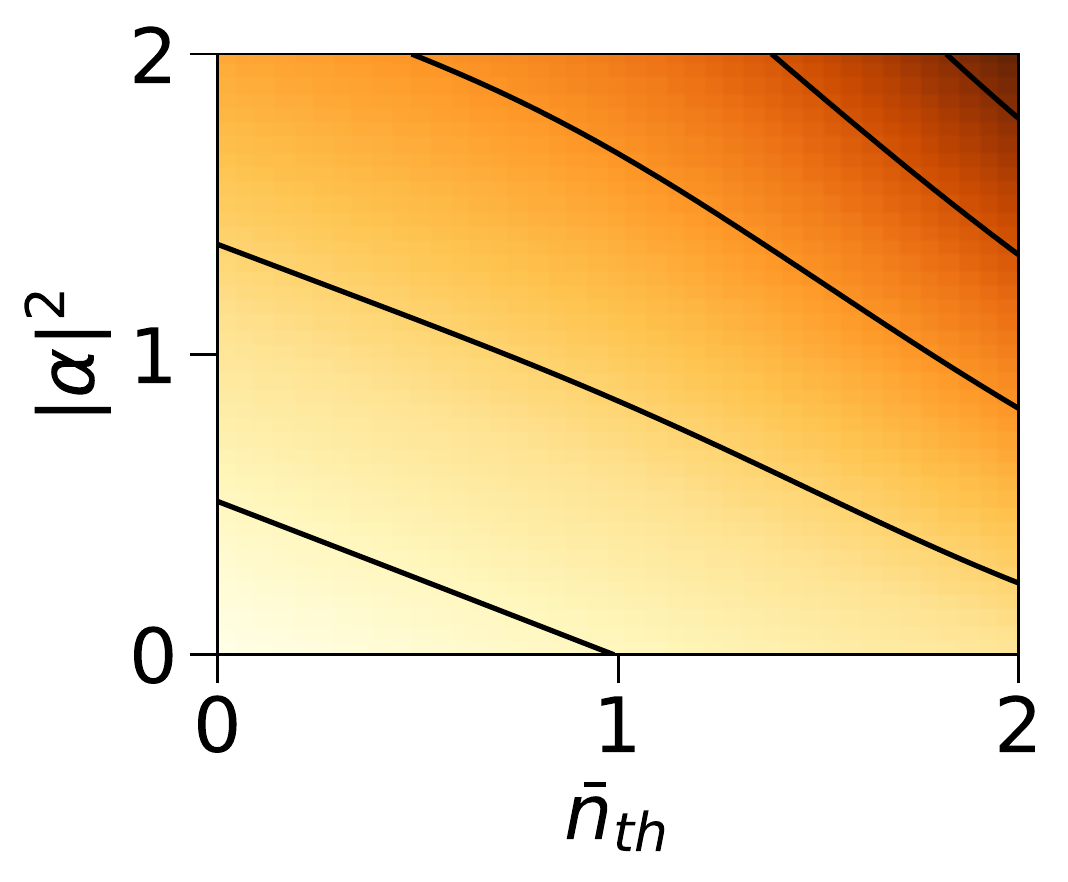}}
\end{subfigure}
\begin{subfigure}[b]{0.239\textwidth}
\raggedright
\resizebox{1.0\textwidth}{!}{
\includegraphics{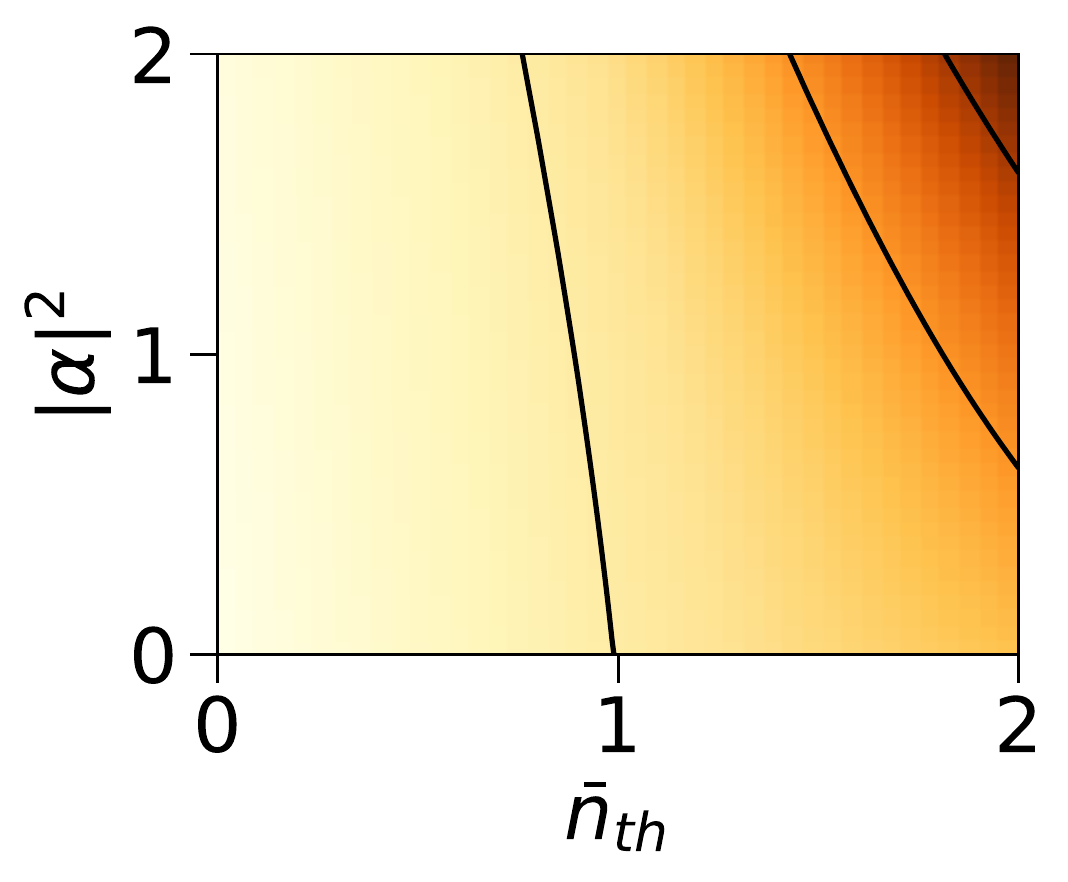}}
\end{subfigure}
\begin{subfigure}[b]{0.239\textwidth}
\raggedright
\resizebox{1.0\textwidth}{!}{
\includegraphics{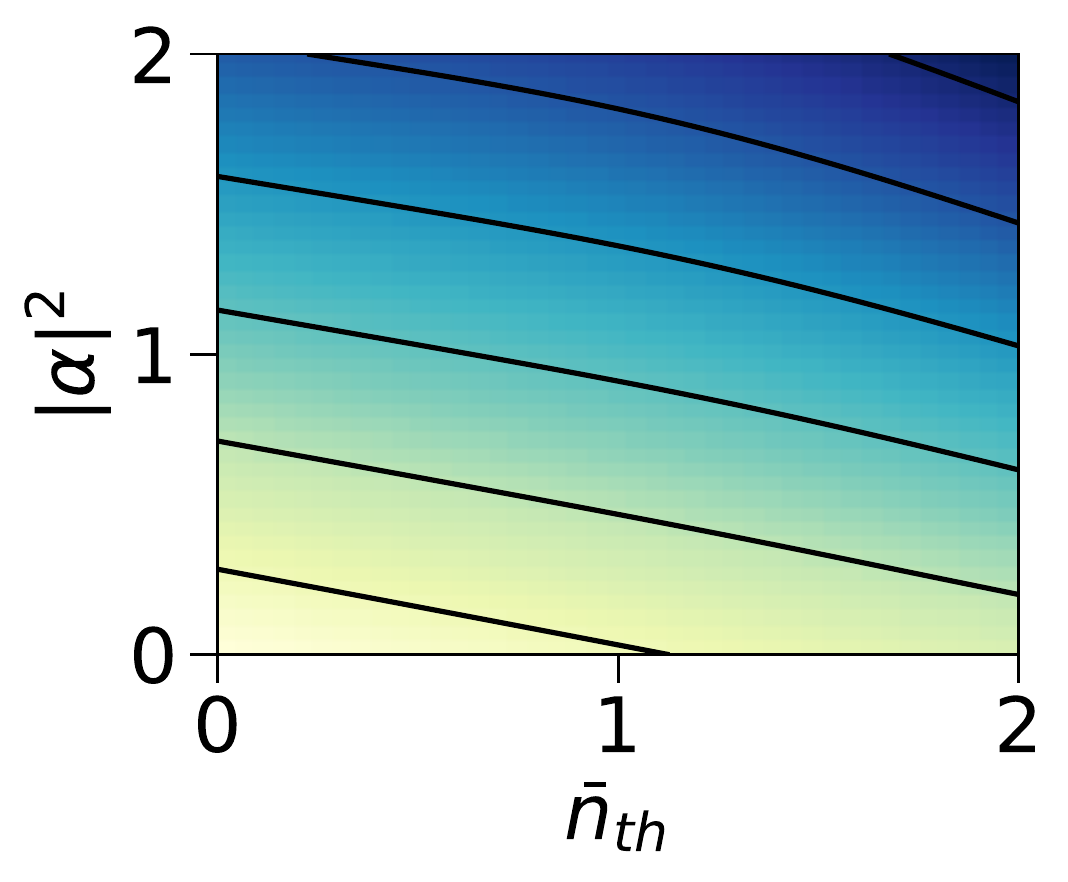}}
\caption{$\phi=0$}
\end{subfigure}
\begin{subfigure}[b]{0.239\textwidth}
\raggedright
\resizebox{1.0\textwidth}{!}{
\includegraphics{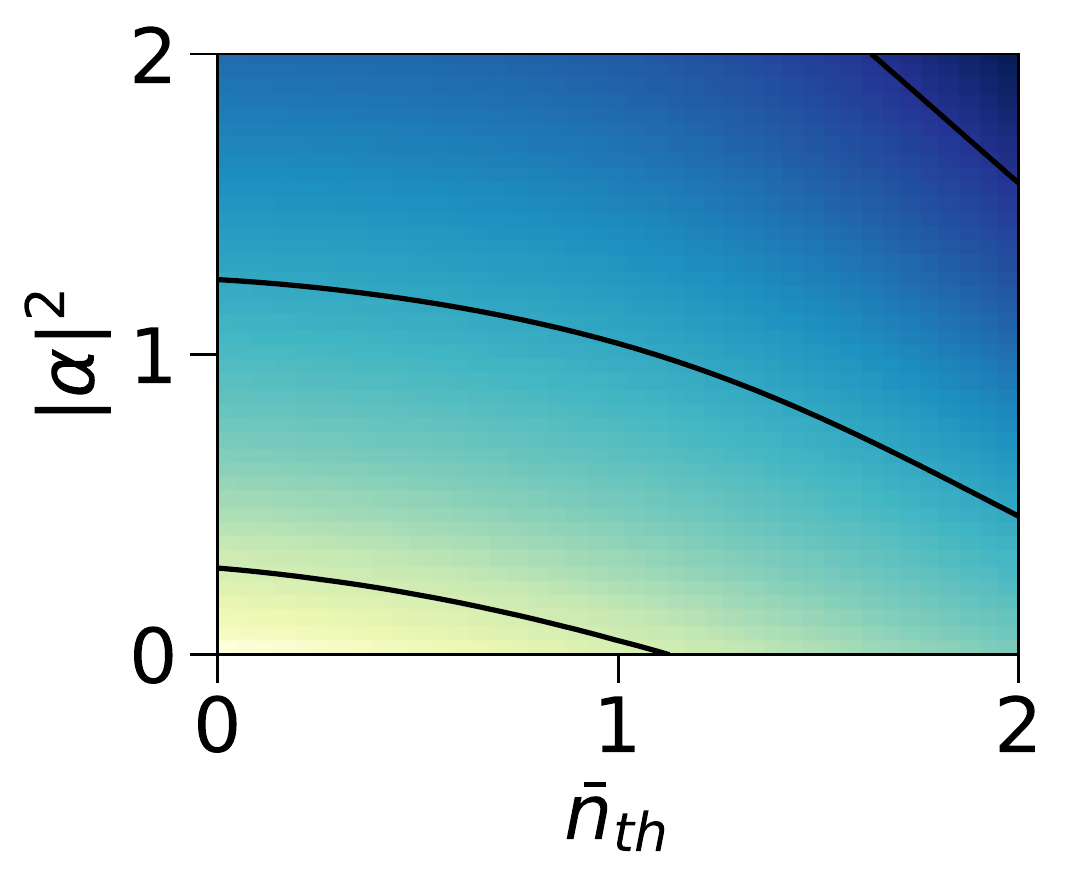}}
\caption{$\phi=\pi/2$}
\end{subfigure}
\caption{\raggedright Infidelity $\mathcal{I}$ (upper plot) and half the diamond distance $\epsilon_\diamond/2$ (lower plot) for an MS gate averaged over a Gaussian distribution of trap frequencies with width $\sigma/2\pi = 600$ Hz and centered at $\nu_0/2\pi=3$ MHz vs.\ $|\alpha|^2$ and $\bar{n}_\text{th}$, for (a) $\phi=0$ and for (b) $\phi=\pi/2$. In each plot, the color represents increasing gate error from lightest to darkest, and the contours start at $0.01$ and increase in steps of $0.01$ from the lower-left to the upper-right.}
\label{fig:ave_fid_contour}
\end{figure}

To better characterize the balance between coherent displacement and thermal excitation prior to the gate, Fig.~\ref{fig:ave_fid_contour} shows how the infidelity $\mathcal{I}$ and half the diamond distance $\epsilon_\diamond/2$ of an MS gate averaged over a Gaussian distribution of trap frequencies with width $\sigma/2\pi = 600$ Hz and centered at $\nu_0/2\pi=3$ MHz depend on $|\alpha|^2$ and $\bar{n}_\text{th}$. The gradient in these plots indicates that increasing either $|\alpha|^2$ or $\bar{n}_\text{th}$ leads to a higher gate error for all initial states. 
For $\phi=0$, the gradient is larger in the $|\alpha|^2$-direction than in the $\bar{n}_\text{th}$-direction, indicating that a coherent displacement prior to the gate is more detrimental than thermal excitation of the same average energy to gate performance. It is therefore worthwhile to seek transport solutions which reduce the amount of coherent displacement, even at the expense of additional thermal excitation due to longer transport times, when $\phi=0$. 

However, when $\phi=\pi/2$, the gradient of $\mathcal{I}$ is much larger in the $\bar{n}_\text{th}$-direction than in the $|\alpha|^2$-direction, even though the gradient of $\epsilon_\diamond/2$ is still larger in the $|\alpha|^2$-direction. This implies that the optimal transport solution depends on the application. One can increase the speed of transport to minimize thermal excitation prior to the gate and reduce $\mathcal{I}$, but the trade-off in increased coherent displacement will raise $\epsilon_\diamond$ and degrade the performance of some quantum algorithms.

\section{Transport measurements} 
\label{sec:exp}

To predict realistic magnitudes of MS-gate error due to small amounts of motional excitation, we apply our simulations to experimentally measured motional spectra of excited Fock states after linear transport. This matches a relevant operational scenario for a trapped-ion quantum computer using the QCCD architecture, in which transport is calibrated for low motional excitation but over time background electric fields arise and result in excess motional heating.  We use the excitation from linear shuttling as representative of transport in general, even though other transport operations, like split/join or junction transport, would likely contribute more excitation.

\begin{figure}[t]
\raggedright
\begin{subfigure}[b]{0.239\textwidth}
\raggedright
\resizebox{1.0\textwidth}{!}{
\includegraphics{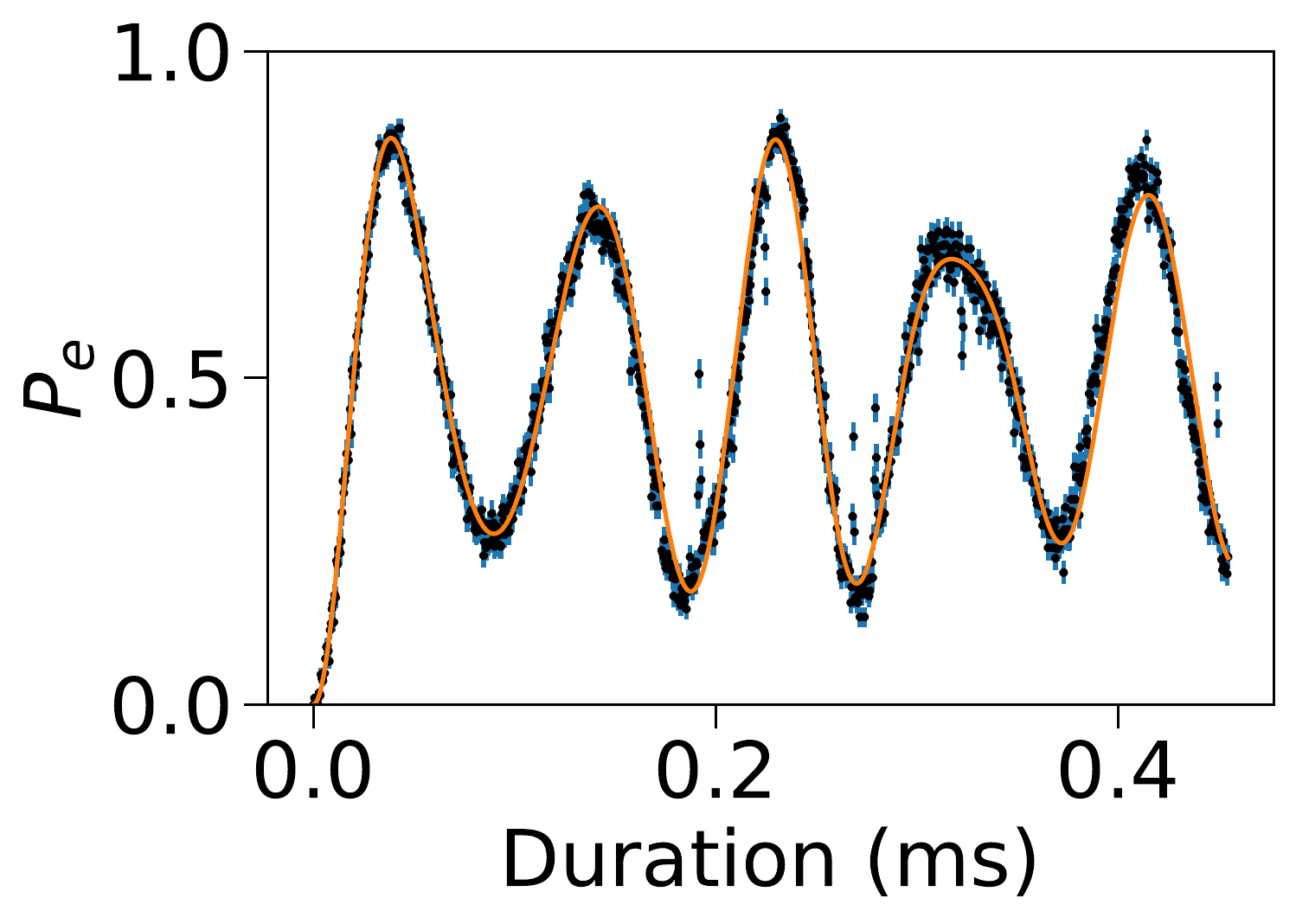}}
\end{subfigure}
\begin{subfigure}[b]{0.239\textwidth}
\raggedright
\resizebox{1.0\textwidth}{!}{
\includegraphics{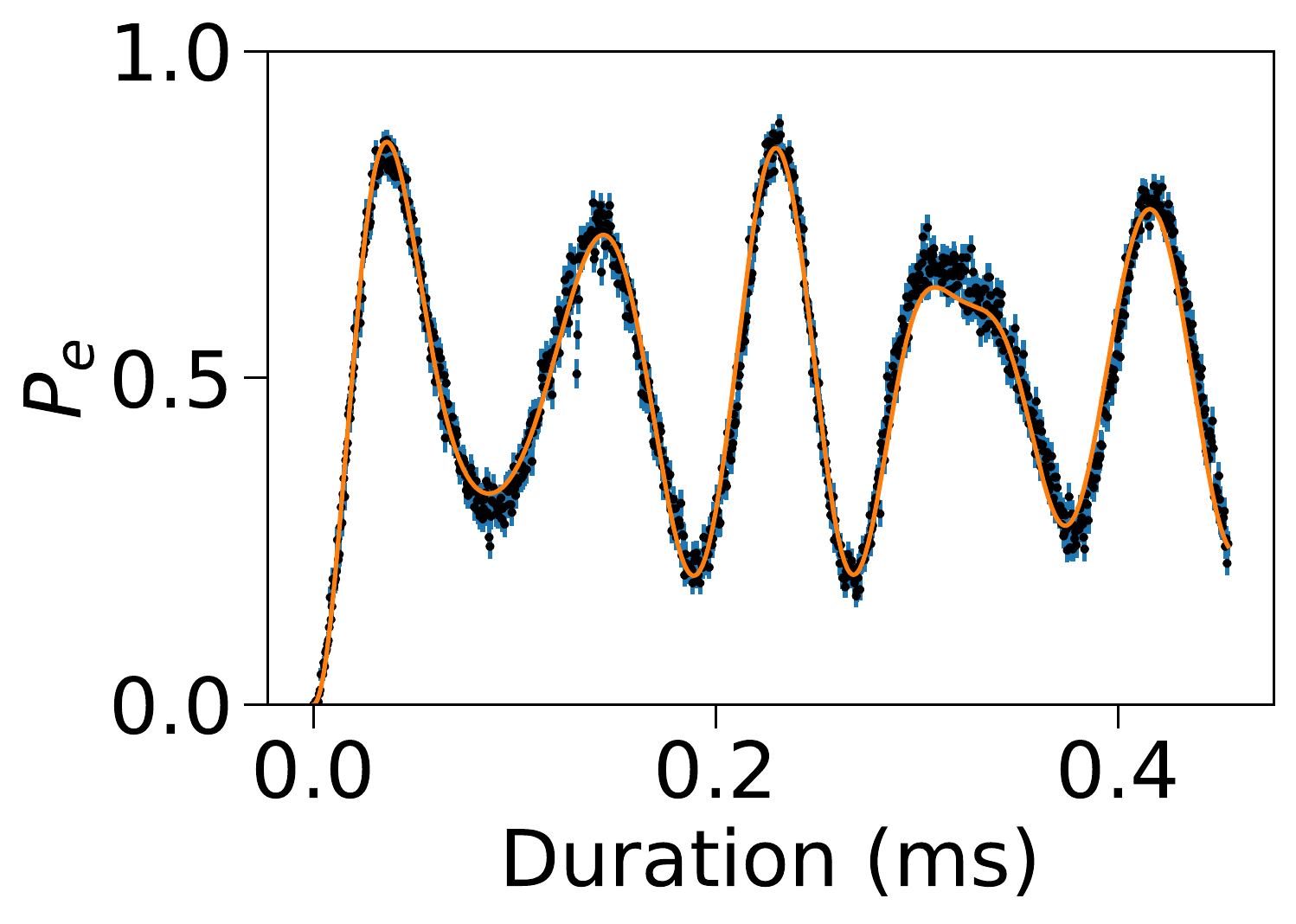}}
\end{subfigure}
\begin{subfigure}[b]{0.239\textwidth}
\raggedright
\resizebox{1.0\textwidth}{!}{
\includegraphics{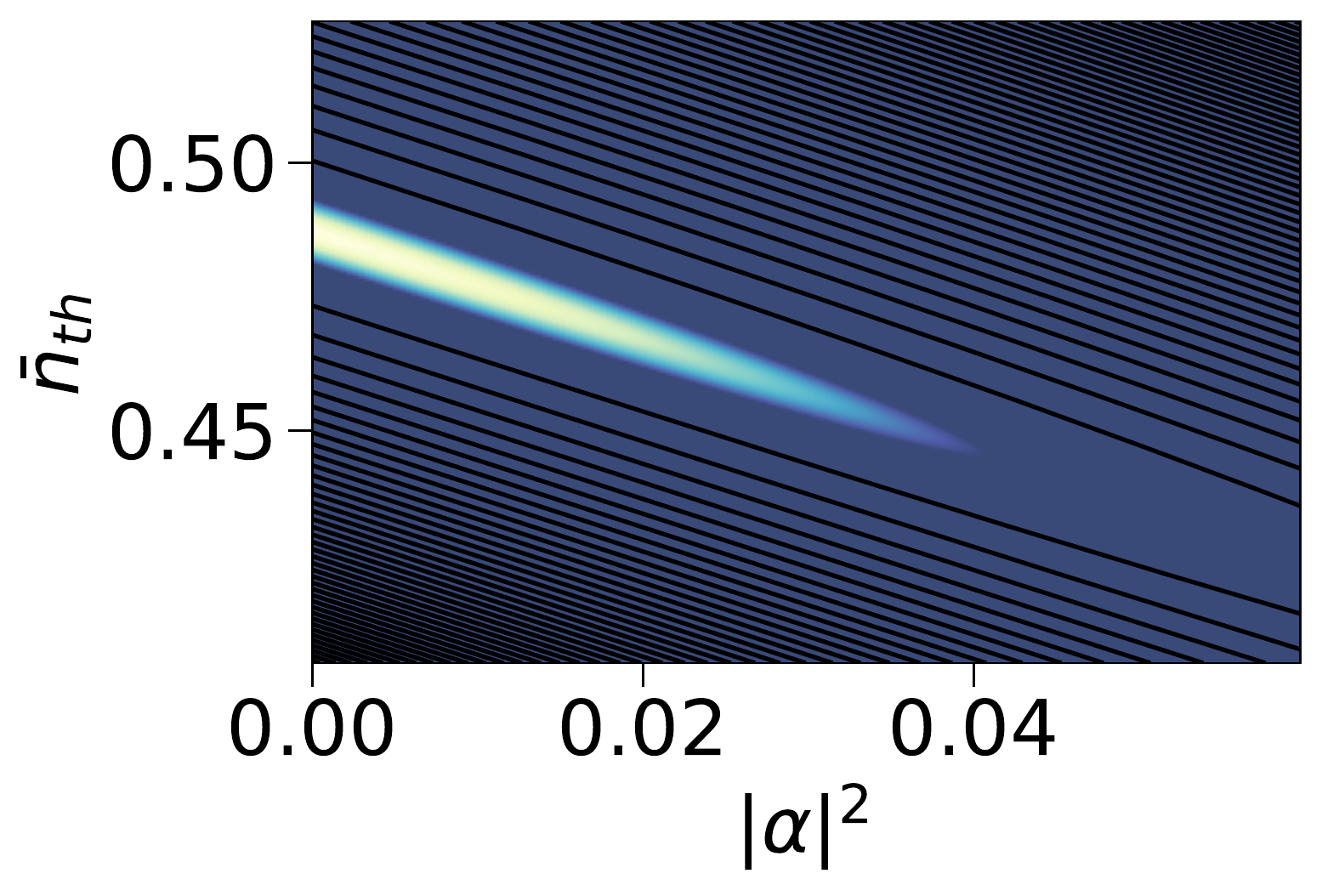}}
\caption{$E_z=0$}
\end{subfigure}
\begin{subfigure}[b]{0.239\textwidth}
\raggedright
\resizebox{1.0\textwidth}{!}{
\includegraphics{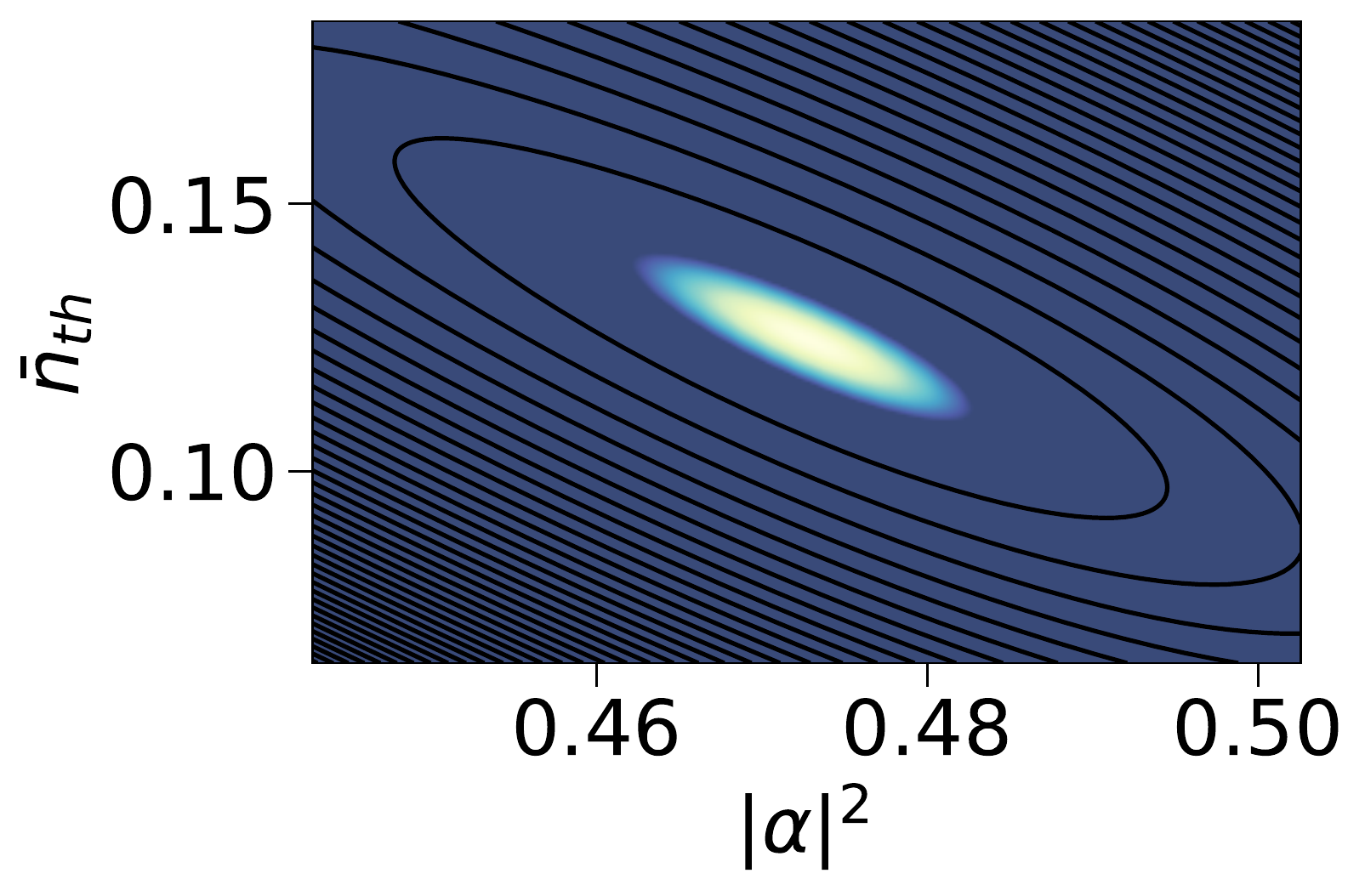}}
\caption{$E_z=40$ V/m}
\end{subfigure}
\caption{\raggedright Excited-state probability $P_e$ during blue-sideband Rabi-flopping experiments and the log-likelihood vs.\ $|\alpha|^2$ and $\bar{n}_\text{th}$ for (a) $E_z=0$ and for (b) $E_z=40$ V/m. In the upper plots, the black dots are the average of $M=500$ measurements at each time step. The blue vertical line segments are the corresponding statistical error bars, and the orange line is the best-fit model of $P_e$ based on a maximum likelihood estimation. In the lower plots, the color scales range from the maximum likelihood (lightest) to $e^{-1}$ times the maximum likelihood (darkest), i.e., to one sigma. The black contours occur every five sigma.}
\label{fig:excited_data}
\end{figure}

In our experiment, the ion is shuttled away and back to its initial position at 16 m/s, and a delay is added at the turn-around point to eliminate most coherent excitation.
After shuttling, we collect blue-sideband Rabi-flopping data to determine the coherent and thermal populations of the transported ion \cite{leibfried:1996}.
Then we apply a controlled electric-field offset of $E_z=40$ V/m in the axial direction to our optimized voltage solution to mimic a background electric field that would typically arise over the course of hours in an experiment, and we collect new blue-sideband Rabi-flopping data. Fig. \ref{fig:excited_data} shows the experimental data, where each data point is an average of $M=500$ shots in the experiment. The error bars shown in the figure represent the statistical uncertainty $\sqrt{P_e(1-P_e)/M}$ of sampling from a binomial distribution, where $P_e$ is the excited-state probability.

We model the blue-sideband Rabi-flopping experiments by assuming ideal Rabi oscillations, except for the addition of 
a phenomenological decoherence rate $\gamma_n = \gamma_0 (n+1)$ between $\ket{n}$ and $\ket{n+1}$. For this model, the excited-state probability $P_e$ during the experiment has the form \cite{leibfried:1997},
\begin{equation}
 P_e = \frac{1}{2} - \frac{1}{2}\sum_{n=0}^\infty P_n \cos(2\Omega_{n,n+1} t) e^{-\gamma_n t},
\end{equation}
where $\Omega_{n,n+1} = \eta \Omega \sqrt{n+1}$ and $P_n=\text{Tr}(\rho_\text{motion}\ket{n}\bra{n})$.

Using the Rabi-flopping data shown in Fig.~\ref{fig:excited_data} for both $E_z=0$ and $E_z=40$ V/m, we perform a maximum likelihood estimation of the model parameters $\Omega$, $\gamma_0$, $|\alpha|^2_0$, $\bar{n}_{\text{th},0}$, $|\alpha|^2_{40}$, and $\bar{n}_{\text{th},40}$. The additional subscript on $|\alpha|^2$ and $\bar{n}_\text{th}$ denotes the value of $E_z$ in V/m, and we demand that the parameters $\Omega$ and $\gamma_0$ are independent of $E_z$. Fig.~\ref{fig:excited_data} shows the values of $P_e$ produced by the best-fit model. The relatively small number of outlying data points, which lie outside the statistical uncertainty of neighboring data points, has a negligible effect on the maximum likelihood estimation. We attribute the cause of the outlying data points to collisions or other catastrophic events that are not captured by the model. 

The estimators for the model parameters are $\Omega/2\pi=136$~kHz, $1/\gamma_0=1.34$ ms, $|\alpha|^2_0=0.00 \pm 0.04$, $\bar{n}_{\text{th},0}=0.49 \pm 0.05$, $|\alpha|^2_{40}=0.47 \pm 0.01$, and $\bar{n}_{\text{th},40}=0.12 \pm 0.02$. 
We have determined the uncertainties by calculating the likelihood for the case of $E_z=0$ and $E_z=40$ V/m, separately, with $\Omega$ and $\gamma_0$ fixed at their optimal values. Fig.~\ref{fig:excited_data} shows contour plots of the log-likelihood vs.\ $|\alpha|^2$ and $\bar{n}_\text{th}$.
For $E_z=40$ V/m, we define the uncertainty in each parameter to be half its maximum range on the curve defined by $e^{-1}$ times the maximum likelihood. For $E_z=0$,
we define the uncertainty in each parameter to be its full range on this curve, noting that $|\alpha|^2_0$ is positive definite and that $\bar{n}_{\text{th},0}$ is highly unlikely to be this much greater than its optimal value.



We then use the estimators of $|\alpha|^2$ and $\bar{n}_{\text{th}}$ to predict the MS-gate error after transport. To represent the conditions of modern ion surface traps, we assume a Gaussian distribution of trap frequencies with width $\sigma/2\pi = 600$ Hz and centered at the optimal value of $\nu_0/2\pi=3$~MHz.
When $E_z=0$, our simulations predict $\mathcal{I}=0.0070$ and $\epsilon_\diamond=0.012$. These values are independent of $\phi$ because $|\alpha|^2_0=0$. When $E_z=40$ V/m, our simulations predict $\mathcal{I}=0.010$ (0.0049) and $\epsilon_\diamond=0.030$ (0.026) for $\phi=0$ $(\pi/2)$.

In this example, we see that a small background electric field of $E_z=40$ V/m raises the gate error by 49\% (150\%) when $\phi=0$, as quantified by $\mathcal{I}$ ($\epsilon_\diamond$), even though the field has decreased $\bar{n}_\text{th}$ by 0.37 and has increased $|\alpha|^2$ by only 0.47, highlighting the sensitivity of MS-gate error to $|\alpha|^2$. When $\phi=\pi/2$, the 40~V/m field reduces $\mathcal{I}$ by 29\% due to the decrease in $\bar{n}_\text{th}$ and the relative insensitivity of $\mathcal{I}$ to $|\alpha|^2$. The field raises $\epsilon_\diamond$ by 110\%, which is 40\% less than when $\phi=0$. This example demonstrates the benefits of optimizing $\phi$ after an experimental implementation of linear ion transport.
Other types of transport are likely to cause greater magnitudes of coherent displacement for the same background electric field, further elevating the importance of reducing coherent displacement and optimizing $\phi$ after transport.

 
 

\section{Conclusion}
\label{sec:conclusion}
We have extended MS-gate models to include both coherent and thermal excitation of motional modes prior to the gate. We have demonstrated that small coherent displacements have a large impact on gate performance and generate significant coherent gate error, making this error source particularly detrimental to quantum algorithms that involve many gates and/or significant ion transport.
Our simulations have focused on Gaussian-distributed trap frequency noise to provide a concrete example, but the interplay between coherent displacement and thermal excitation is important for a broad set of experimental realities with a diverse spectrum of both environmental and control-based noise sources.
We have also validated our model of ion motion against measurements of the motional distribution after linear transport, and we have applied our simulations to predict MS-gate performance in a realistic experimental situation. As trapped-ion quantum processors scale up to larger numbers of qubits and support next-generation quantum algorithms, the analysis and methods presented in this paper will help maximize performance
by assessing the trade-offs between operations that produce coherent and incoherent excitation of ion motion, a paradigm that is relevant to other quantum-computing technologies and motional quantum sensors.

\section*{Acknowledgments}
We thank Kevin Young for fruitful discussions. This research was funded by the U.S. Department of Energy, Office of Science, Office of Advanced Scientific Computing Research.
Sandia National Laboratories is a multimission laboratory managed and operated by National Technology \& Engineering Solutions of Sandia, LLC, a wholly owned subsidiary of Honeywell International Inc., for the U.S. Department of Energy's National Nuclear Security Administration under contract DE-NA0003525.  This paper describes objective technical results and analysis. Any subjective views or opinions that might be expressed in the paper do not necessarily represent the views of the U.S. Department of Energy or the United States Government. 

\section*{References}
\bibliography{ion_refs}

\end{document}